\documentclass[
reprint,
superscriptaddress,
aps,
 amsmath,amssymb,
 prl,
]{revtex4-2}

\usepackage{graphicx}
\usepackage{dcolumn}
\usepackage{bm}
\usepackage{upgreek}
\usepackage{siunitx}
\usepackage[most]{tcolorbox}
\definecolor{IOPBlue}{HTML}{1E8FFF}
\tcbset{colback=white, colframe=IOPBlue, arc=0mm, left=0mm, right=0mm,top=0.5mm,bottom=.5mm,
        highlight math style= {enhanced, 
            colframe=red,colback=red!10!white,boxsep=0pt}}
\usepackage{dsfont}
\usepackage{orcidlink}

\usepackage[mathlines]{lineno}

\begin{document}

\preprint{APS/123-QED}

\newcommand{\berlin}{Institut f\"{u}r Physik, Humboldt-Universit\"{a}t zu Berlin, Newtonstr.~~15, 12489 Berlin, Germany}
\newcommand{\bonn}{Helmholtz-Institut f\"{u}r Strahlen- und Kernphysik, Rheinische Friedrich-Wilhelms-Universit\"{a}t Bonn, Nussallee 14-16, 53115 Bonn, Germany}
\newcommand{\chulalongkorn}{Department of Physics, Faculty of Science, Chulalongkorn University, Bangkok 10330, Thailand}
\newcommand{\cmu}{Department of Physics, Carnegie Mellon University, Pittsburgh, PA 15213, USA}
\newcommand{\etp}{Institute of Experimental Particle Physics~(ETP), Karlsruhe Institute of Technology~(KIT), Wolfgang-Gaede-Str.~1, 76131 Karlsruhe, Germany}
%

\newcommand{\iap}{Institute for Astroparticle Physics~(IAP), Karlsruhe Institute of Technology~(KIT), Hermann-von-Helmholtz-Platz 1, 76344 Eggenstein-Leopoldshafen, Germany}
\newcommand{\ipe}{Institute for Data Processing and Electronics~(IPE), Karlsruhe Institute of Technology~(KIT), Hermann-von-Helmholtz-Platz 1, 76344 Eggenstein-Leopoldshafen, Germany}
\newcommand{\itep}{Institute for Technical Physics~(ITEP), Karlsruhe Institute of Technology~(KIT), Hermann-von-Helmholtz-Platz 1, 76344 Eggenstein-Leopoldshafen, Germany}
%
%
%

\newcommand{\infnbicocca}{Istituto Nazionale di Fisica Nucleare (INFN) -- Sezione di Milano-Bicocca, Piazza della Scienza 3, 20126 Milano, Italy}

\newcommand{\infnmilano}{Istituto Nazionale di Fisica Nucleare (INFN) -- Sezione di Milano, Via Celoria 16, 20133 Milano, Italy}

\newcommand{\polimi}{Politecnico di Milano, Dipartimento di Elettronica, Informazione e Bioingegneria, Piazza L. da Vinci 32, 20133 Milano, Italy}

\newcommand{\umilano}{Dipartimento di Fisica, Universit\`{a} di Milano - Bicocca, Piazza della Scienza 3, 20126 Milano, Italy}

\newcommand{\inr}{Institute for Nuclear Research of Russian Academy of Sciences, 60th October Anniversary Prospect 7a, 117312 Moscow, Russia}
\newcommand{\inrfootnote}{Institutional status in the KATRIN Collaboration has been suspended since February 24, 2022}
\newcommand{\lbnl}{Nuclear Science Division, Lawrence Berkeley National Laboratory, Berkeley, CA 94720, USA}
\newcommand{\madrid}{Departamento de Qu\'{i}mica F\'{i}sica Aplicada, Universidad Autonoma de Madrid, Campus de Cantoblanco, 28049 Madrid, Spain}
\newcommand{\mainz}{Institut f\"{u}r Physik, Johannes-Gutenberg-Universit\"{a}t Mainz, 55099 Mainz, Germany}

\newcommand{\mpp}{Max Planck Institute for Physics, Boltzmannstr. 8, 85748 Garching, Germany}

\newcommand{\massit}{Laboratory for Nuclear Science, Massachusetts Institute of Technology, 77 Massachusetts Ave, Cambridge, MA 02139, USA}
\newcommand{\muenster}{Institute for Nuclear Physics, University of M\"{u}nster, Wilhelm-Klemm-Str.~9, 48149 M\"{u}nster, Germany}
\newcommand{\npi}{Nuclear Physics Institute,  Czech Academy of Sciences, 25068 \v{R}e\v{z}, Czech Republic}
\newcommand{\unc}{Department of Physics and Astronomy, University of North Carolina, Chapel Hill, NC 27599, USA}
\newcommand{\washington}{Center for Experimental Nuclear Physics and Astrophysics, and Dept.~of Physics, University of Washington, Seattle, WA 98195, USA}
\newcommand{\wuppertal}{Department of Physics, Faculty of Mathematics and Natural Sciences, University of Wuppertal, Gau{\ss}str.~20, 42119 Wuppertal, Germany}
\newcommand{\saclay}{IRFU (DPhP \& APC), CEA, Universit\'{e} Paris-Saclay, 91191 Gif-sur-Yvette, France }

\newcommand{\suranaree}{School of Physics and Center of Excellence in High Energy Physics and Astrophysics, Suranaree University of Technology, Nakhon Ratchasima 30000, Thailand}
\newcommand{\tum}{Technical University of Munich, TUM School of Natural Sciences, Physics Department, James-Franck-Stra\ss e 1, 85748 Garching, Germany}
\newcommand{\uhd}{Institute for Theoretical Astrophysics, University of Heidelberg, Albert-Ueberle-Str.~2, 69120 Heidelberg, Germany}
\newcommand{\tunl}{Triangle Universities Nuclear Laboratory, Durham, NC 27708, USA}
%
%
\newcommand{\ornl}{Also affiliated with Oak Ridge National Laboratory, Oak Ridge, TN 37831, USA}
%

\affiliation{\iap}
\affiliation{\ipe}
\affiliation{\muenster}
\affiliation{\infnbicocca}
\affiliation{\polimi}
\affiliation{\infnmilano}
\affiliation{\chulalongkorn}
\affiliation{\cmu}
\affiliation{\etp}
\affiliation{\madrid}
\affiliation{\washington}
\affiliation{\npi}
\affiliation{\tum}
\affiliation{\mpp}
\affiliation{\wuppertal}
\affiliation{\massit}
\affiliation{\unc}
\affiliation{\tunl}
\affiliation{\suranaree}
\affiliation{\saclay}
\affiliation{\lbnl}
\affiliation{\umilano}
\affiliation{\itep}
\affiliation{\berlin}
\affiliation{\uhd}
\affiliation{\mainz}
\affiliation{\inr}



\title{First constraints on general neutrino interactions based on KATRIN data}


\author{M.~Aker}\affiliation{\iap}
\author{D.~Batzler}\affiliation{\iap}
\author{A.~Beglarian}\affiliation{\ipe}
\author{J.~Beisenk\"{o}tter}\affiliation{\muenster}
\author{M.~Biassoni}\affiliation{\infnbicocca}
\author{B.~Bieringer}\affiliation{\muenster}
\author{Y.~Biondi}\affiliation{\iap}
\author{F.~Block}\affiliation{\iap}
\author{B.~Bornschein}\affiliation{\iap}
\author{L.~Bornschein}\affiliation{\iap}
% ...

\collaboration{KATRIN Collaboration}

\begin{abstract}
The precision measurement of the tritium $\upbeta$-decay spectrum performed by the KATRIN experiment provides a unique way to search for general neutrino interactions (GNI). All theoretical allowed GNI terms involving neutrinos are incorporated into a low-energy effective field theory, and can be identified by specific signatures in the measured tritium $\upbeta$-spectrum. In this paper an effective description of the impact of GNI on the $\upbeta$-spectrum is formulated and the first constraints on the effective GNI
parameters are derived based on the 4 million electrons collected in the second measurement campaign of KATRIN in 2019. 
In addition, constraints on selected types of interactions are investigated, thereby exploring the potential of KATRIN to search for more specific new physics cases, including a right-handed W boson, a charged Higgs or leptoquarks.
\end{abstract}

\keywords{Neutrino, General Neutrino Interactions}
\maketitle


\textit{Introduction} \----
Neutrinos are considered a prime window to new physics, with a non-zero neutrino mass calling for an extension of the Standard Model of particle physics already.
The nature of these new physics contributions remains unclear due to the abundance of theories and the elusive nature of the neutrino.
A particularly broad and therefore powerful approach to identify such new physics is offered by the theory of general neutrino interactions (GNI) \cite{Ludl2016DirectInteractions,Bischer2019GeneralPerspective,Bischer2021EffectivePhenomenology}, which describes the weak interaction beyond its Standard Model V-A structure in a more generalized manner than the already well-studied neutrino non-standard interactions (NSI) \cite{Dev2019NeutrinoReport,Ohlsson2013StatusInteractions,Farzan2018NeutrinoInteractions,Coloma2023GlobalElectrons,Amaral2023ALandscape}. The new low-energy effective interaction terms of the GNI theory can also be generated by (sterile-neutrino extended) Standard Model effective field theory (SM(N)EFT) operators of mass dimension 6 \cite{Wise2014EffectiveInteractions,Han2020ScalarInteractions,Cirigliano2013BetaEra,Bischer2019GeneralPerspective}. 
GNI therefore connect new physics phenomena both below and above the weak scale through the effective field theory (EFT). 
In recent years a multitude of investigations have been conducted for various neutrino interaction channels, reaching from neutrino-oscillation experiments \cite{Bischer2019GeneralDetector,Khan2020BorexinoInteractions,Gupta2023NeutrinoNSI} to neutrino-scattering processes \cite{Li2020GeneralDecays,Chen2021ConstraintsExperiments,AristizabalSierra2018COHERENTInteractions,Papoulias2018COHERENTPhysics,Rodejohann2017DistinguishingInteractions,Lindner2017CoherentInteractions} to studies of $\upbeta$-decay processes \cite{KumarBanerjee2023TestingPTOLEMY,Gonzalez-Alonso2018NewDecay,Cirigliano2013BetaEra,Han2020ScalarInteractions}. 
In this work, we examine the impact of GNI on the $\upbeta$-decay spectrum of tritium as measured by KATRIN, also considering the potential existence of additional eV-scale neutrino-mass states. \\

\textit{Theoretical Formalism} \----
The GNI theory includes all possible Lorentz-invariant operators for four-fermion interactions involving at least one neutrino. With this approach, the GNI Lagrangian for charged-current interactions can be expressed as

\begin{eqnarray}\label{equ:CC}
    \mathcal{L}_{\mathrm{GNI}}^{\mathrm{CC}}= & -\frac{G_{\mathrm{F}} V_{\gamma\delta}}{\sqrt{2}} \sum_{j=1}^{10}\left(\overset{(\sim)}{\epsilon}_{j,\mathrm{ud}}\right)^{\alpha\beta\gamma\delta}\nonumber\\
    & \cdot\left( \bar{\mathrm{e}}_{\alpha}\mathcal{O}_j \nu_{\beta}\right) \left( \bar{\mathrm{u}}_{\gamma}\mathcal{O}'_j \mathrm{d}_{\delta}\right) + \mathrm{h.c.},
\end{eqnarray}
where $\mathcal{O}^{(')}_j$ are operators describing each new kind of interaction $j$, see table~\ref{tab:GNIOperator} in the appendix. $\epsilon_j$, $\tilde{\epsilon}_{j}$ are the flavor-space tensors, expressing the strength of interaction $j$ with respect to the Standard Model Fermi interaction. $V_{\gamma\delta}$ denotes the CKM matrix and $G_\mathrm{F}$ is the Fermi constant. Greek indices run over flavor. 

As no assumption is made regarding the mechanism generating the neutrino mass, heavier neutrinos with masses up to the mass of the W boson are also accounted for in the theory. Similarly, the theory imposes no assumptions regarding the Dirac or Majorana nature of neutrinos.

The mapping of low-energy GNI terms onto dimension-6 SM(N)EFT operators allows for the description of new physics above the weak scale. Assuming a common high-energy origin of new physics allows to set strong indirect constraints on the GNI couplings from existing constraints on the SM(N)EFT couplings. 
Low-energy GNI contributions without a matching counterpart at dimension 6
may indicate the existence of low-energy new physics or the necessity of expanding the EFT to a higher order \cite{Bischer2019GeneralPerspective}.\\

\textit{The KATRIN Experiment} \----
The KArlsruhe TRItium Neutrino experiment (KATRIN) performs a direct, kinematic measurement of the effective mass of the electron antineutrino in the endpoint region of the $\upbeta$-decay spectrum of tritium. KATRIN features a \SI{e11}{Bq} high-activity windowless gaseous molecular tritium source, where the $\upbeta$-electrons are emitted and are magnetically guided through the setup \cite{Design2021} towards the detector. In the KATRIN spectrometer, the electrons' momenta are magnetically collimated and the electrons are electrostatically
filtered based on their kinetic energy with a filter width of $\mathcal{O}(\SI{1}{eV})$ \cite{Beamson1980TheSpectrometer,Lobashev1985AMass,Picard1992AElectrons}.
The filter threshold is varied in steps of $\mathcal{O}(\SI{1}{eV})$ close to the spectral endpoint $E_0$ in order to record an integral $\upbeta$-spectrum. 
The KATRIN collaboration has recently set a new world-leading direct upper limit on the effective electron antineutrino mass to \SI{.45}{eV} at 90\,\% confidence level (CL) \cite{Aker2024DirectData}.
The high level of understanding of the various experimental effects combined with the excellent theoretical knowledge of 
the tritium $\upbeta$-decay spectrum makes KATRIN an ideal instrument for precision measurements of new physics phenomena.
Unlike the neutrino mass, showing a pronounced signature close to the endpoint, sterile neutrinos \cite{Aker2022ImprovedCampaign,Siegmann2024DevelopmentExperiment} and general neutrino interactions can impact the shape of the entire recorded spectrum.\\

\textit{Second Measurement Campaign} \----
The analysis presented here is based on the data of the second measurement campaign (KNM2) of KATRIN, which was conducted from October to December 2019 with about 45 measurement days. As in the neutrino-mass analysis, the data recorded at the scan steps within the range $[E_0-\SI{40}{eV},\, E_0+\SI{135}{eV}]$ is used.
The KNM2 campaign was selected for the proof-of-principle study at hand because it features a background reduced by 25\,\% to \SI{0.22}{cps}, and an increased signal rate due to a 3.9 times higher tritium gas density in the source of \SI{4.2e21}{\per\square\meter}, compared to the first neutrino-mass measurement \cite{Aker2019}. 
Further details on the data set can be found in ref. \cite{Aker2022DirectSensitivity}.\\
 
\begin{figure}
    \centering
    \includegraphics[width=.5\textwidth]{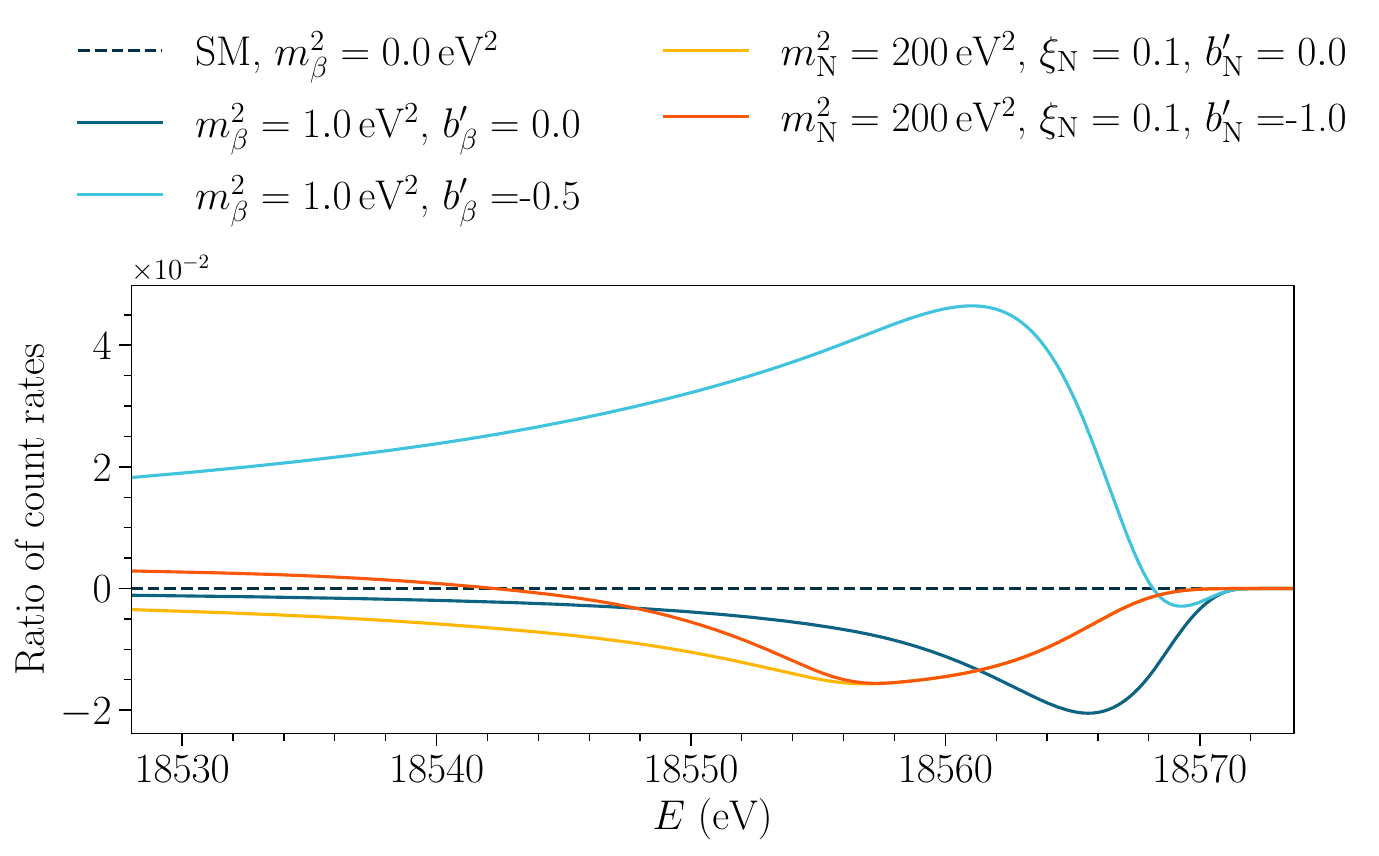}
    \caption{Effect of different GNI contributions on the endpoint region of the integral $\upbeta$-spectrum. Both cases with (N) and without ($\upbeta$) the presence of an additional heavier neutrino have been considered. The position of the kink-like structure in the signal is dependent on the (heavy) neutrino mass $m^2_{\upbeta(\mathrm{N})}$. The endpoint is here at \SI{18573.7}{eV}.}
    \label{fig:GNI-Speck}
\end{figure}

\textit{The Spectral Model} \----
The tritium differential decay rate in the presence of GNI can be derived from the sum of the GNI Lagrangian in equation~\ref{equ:CC} and the Lagrangian describing the Standard Model contribution.
The new kinds of interactions with respect to both light (indicated by index $\upbeta$) and additional heavier neutrinos (index N) are considered in a left-right-symmetric model. 
Further details on the computation are presented in the appendix. The differential decay rate then takes the form:

\begin{eqnarray}\label{equ:totdiffdecrate}
    &\frac{\mathrm{d}\Gamma}{\mathrm{d}E}= \frac{G_{\mathrm{F}}^2 V_{\mathrm{ud}}^2}{2\pi^3} F(E,\,2) \sqrt{(E+m_\mathrm{e})^2-m_\mathrm{e}^2} (E+m_\mathrm{e})\nonumber\\
    & \cdot\sum\limits_j \sum\limits_{k=\upbeta ,\,\mathrm{N}} \zeta_j \varepsilon_j \sqrt{\varepsilon_j^2-m_{k}^2} \xi_k \left[1-  b'_k \frac{m_k}{\varepsilon_j}  \right] \Theta(\varepsilon_j-m_k) 
\end{eqnarray}
The Fermi function, $F(E,\,2)$, models the electromagnetic interaction of the emitted electrons with the daughter nucleus. The electrons' kinetic energy is denoted by $E$ and their mass by $m_\mathrm{e}$. The impact of the molecular final states is considered by incorporating the molecular excitation energy, $V_j$, which reduces the available neutrino energy, $\varepsilon_j=E_0-E-V_j$. The probability of producing the excited state $j$ is expressed by $\zeta_j$. Beyond that, radiative corrections are taken into account, while further theoretical corrections are neglected \cite{Kleesiek2018}.
Both the $\xi_k$ and the $b_k^{\prime}$ parameters are defined in terms of the flavor-space tensors ${\epsilon}_{j,\mathrm{ud}}$ and $\tilde{\epsilon}_{j,\mathrm{ud}}$, expressing the coupling strength of GNI to left- and right-handed neutrinos, respectively, the mixing matrices $U_{\mathrm{e}i}$ and $T_{\mathrm{e}i}$ for the left- and right-handed light neutrino, the mixing matrices $S_{\mathrm{e}i}$ and $V_{\mathrm{e}i}$ for the left- and right-handed heavy neutrino, defined according to \cite{Ludl2016DirectInteractions}, and the nuclear form factors $g_\mathrm{V}$, $g_\mathrm{S}$, $g_\mathrm{T}$, and $g_\mathrm{A}$, see table \ref{tab:couplingconst} in the appendix. The Standard Model differential decay rate can be restored by setting $\xi_{\mathrm{N}}=b_k^{\prime}=0$.
The distinct signatures of $\xi_k$ and $b'_k$ on the $\upbeta$-spectrum are visualized in figure~\ref{fig:GNI-Speck}. The explicit dependence of the effective GNI parameters $\xi_k$ and $b'_k$ on the $\epsilon$ parameters can be found in the appendix. The differential decay rate as defined in equation \ref{equ:totdiffdecrate} is included in the prediction of the observed integral rate via
\begin{equation}
            R(qU_i) = A_\mathrm{s} N_\mathrm{T} \int\displaylimits_{qU_i}^{E_0} \frac{\mathrm{d}\Gamma (E)}{\mathrm{d}E}  f(E,qU_i) \mathrm{d}E + R_\mathrm{bg}(qU_i).
\end{equation}
Here, $f(E,qU_i)$ represents the experimental response function at each retarding-energy set point $qU_i$, which describes the transmission properties of the experiment, as well as accounting for additional physical effects such as electron scatterings. $q$ is the electrons' charge and $U_i$ the electrostatic retarding potential of the spectrometer at scan step $i$.
$N_\mathrm{T}$ serves as a signal renormalization factor for the normalized amplitude $A_\mathrm{s}$, and is based on the number of tritium atoms in the source, the maximum acceptance angle due to the magnetic field design, and the detection efficiency. $R_\mathrm{bg}$ is the background rate.\\

\textit{Analysis Procedure} \----
The presented analysis separately investigates a case with light (quantities with index $\upbeta$) and with additional heavier neutrinos (index N). Consequently, two distinct sets of parameters of interest are considered. In order to infer the parameters of interest 
in the absence of any heavy neutrinos, we focus on the two-dimensional parameter space comprising the effective neutrino mass $m_\upbeta$ and the shape-modifying GNI parameter $b'_\upbeta$.
In the presence of heavier neutrinos, the three-dimensional parameter space is defined by the heavier neutrino-mass state $m_4^2$ and the GNI parameters $\xi_\mathrm{N}$, in units of $\xi_\upbeta=g_\mathrm{V}^2+3g_\mathrm{A}^2$, and $b'_\mathrm{N}$. The amplitude of the additional neutrino branch is scaled by $\xi_\mathrm{N}$, while specific shape deformations are introduced by $b'_\mathrm{N}$, as can be seen in figure~\ref{fig:GNI-Speck} and equation~\ref{equ:totdiffdecrate}. 
$E_0,\; A_\mathrm{s},$ and the constant background component $R_\mathrm{bg}^\mathrm{base}$ are added as nuisance parameters.

The GNI parameter space is probed on a grid, for which the model is fit to the data with respect to the nuisance parameters at each grid point using $\chi^2$-function minimization. This allows for a numerically robust minimization of the $\chi^2$-function.
The propagation of systematic uncertainties is conducted using a pull-term approach \cite{Aker2022DirectSensitivity}. 

In the absence of a heavy neutrino, a $60\times30$ linearly spaced grid over the parameter space $m_\upbeta^2 \times b'_\upbeta$ within the intervals $[\SI{0}{eV^2},\,\SI{2.2}{eV^2}]\times[-1.0,\,1.0]$ has been employed. The three-dimensional parameter space for the case including heavy neutrinos is defined by $30\times40$ logarithmically spaced values of the parameters $m_\mathrm{N}^2\times \xi_\mathrm{N}/\xi_\upbeta$ within $[\SI{1}{eV^2},\,\SI{1600}{eV^2}]\times[\num[exponent-product=\ensuremath{\cdot}]{3.6e-4},\,10.0]$ times 9 linearly spaced values for $b'_\mathrm{N}$ within $[-1.0,\,1.0]$.
The exclusion contours at the 95\,\% CL are drawn at $\Delta\chi^2_\upbeta=3.84$ and $\Delta\chi^2_\mathrm{N}=5.99$ for the grids without and with heavy neutrinos. 
The critical values of the $\Delta\chi^2$ distribution have been derived through Monte Carlo simulations on $\mathcal{O}(\num{10000})$ randomized spectra. 

The correctness of the implementation of the model has been confirmed by the conformity of the results with 
those obtained with the independent sterile-neutrino analysis framework at KATRIN, see figure~8 in \cite{Aker2022ImprovedCampaign}. %
Furthermore, to mitigate human-induced biases, a blinding scheme was applied. The analysis procedure was first applied to an Asimov data set. Next, a simulated data set with statistical fluctuations was used to test the robustness of the procedure. Thereafter, the analysis procedure was fixed and applied to the experimental data.\\

\textit{Results} \----
When applying this analysis procedure to the KNM2 data set, no significant indication of a GNI-like spectral distortion could be identified in either case, with or without an additional heavy neutrino-mass state. Also, the uncertainties in both cases are dominated by statistics, as shown in figures~\ref{fig:SysBreaksownSM} and \ref{fig:SysBreakdownST} in the appendix. Figure~\ref{fig:SM-contour} presents the 95\,\% CL exclusion contour for the GNI; 
The region with high values of $m_\upbeta^2$ is excluded. Furthermore, a correlation between $m_\upbeta^2$ and $b'_\upbeta$ can be observed, which makes it more challenging to constrain the latter. Consequently, the effect of additional charged currents on the neutrino-mass observable is slight \cite{Bonn2011TheDecay}.
The best fit is located at $b'_\upbeta=-1$ and $m_\upbeta^2=\SI{0.11}{eV^2}$ with a significance with respect to the null hypothesis of $\Delta\chi^2=\chi^2_\mathrm{NH}-\chi^2_\mathrm{min}=0.003$, thus implying no statistical significance. A combination of external constraints suggests $|b'|<0.26$ (95\,\% CL) \cite{Bonn2011TheDecay}. Furthermore, the projected final sensitivity for KATRIN after \num{1000} days measurement time was estimated using an Asimov data set assuming $m_\upbeta^2=\SI{0}{eV^2}$. It is apparent that the sensitivity to $m_\upbeta^2$ is considerably enhanced while the impact of $b'_\upbeta$ is diminished for small neutrino masses.

The exclusion contours at 95\,\% CL for the GNI acting on an additional heavier neutrino-mass state are displayed in figure~\ref{fig:ST-contour}. In this primary analysis, the value of  $m_\upbeta^2$ was fixed to \SI{0}{eV^2}. 
The sensitivity contours visibly depend on $b'_\mathrm{N}$: It enhances the kink-like signature due to the heavier mass state for negative values, whereas it diminishes the manifestation of the signal structure for more positive values.
Moreover, the sensitivity is markedly enhanced in the higher-mass region. This will become relevant when KATRIN commences its search for keV sterile neutrinos with the planned TRISTAN detector upgrade \cite{Barry2014SterileKATRIN,Siegmann2024DevelopmentExperiment}. The best fit is found at $m_\mathrm{N}^2=\SI{97.45}{eV^2}$, $\xi_\mathrm{N}/\xi_\upbeta=0.005$, and $b'_\mathrm{N}=-1$. The significance of the best fit compared to the null hypothesis is $\Delta\chi^2=3.338$ and is thus not statistically significant.
Furthermore, the estimate of the projected final sensitivity corresponding to \num{1000} days measurement time at KATRIN demonstrates a substantial improvement of approximately one order of magnitude in terms of $\xi_\mathrm{N}/\xi_\upbeta$. In addition, there is potential for a further order-of-magnitude improvement by eliminating parameter correlations with respect to $E_0$ through a more precise understanding of the absolute energy scale in KATRIN \cite{Steinbrink2017StatisticalKATRIN,Rodenbeck2022AExperiment}.\\

\begin{figure}
    \centering
    \includegraphics[width=.5\textwidth]{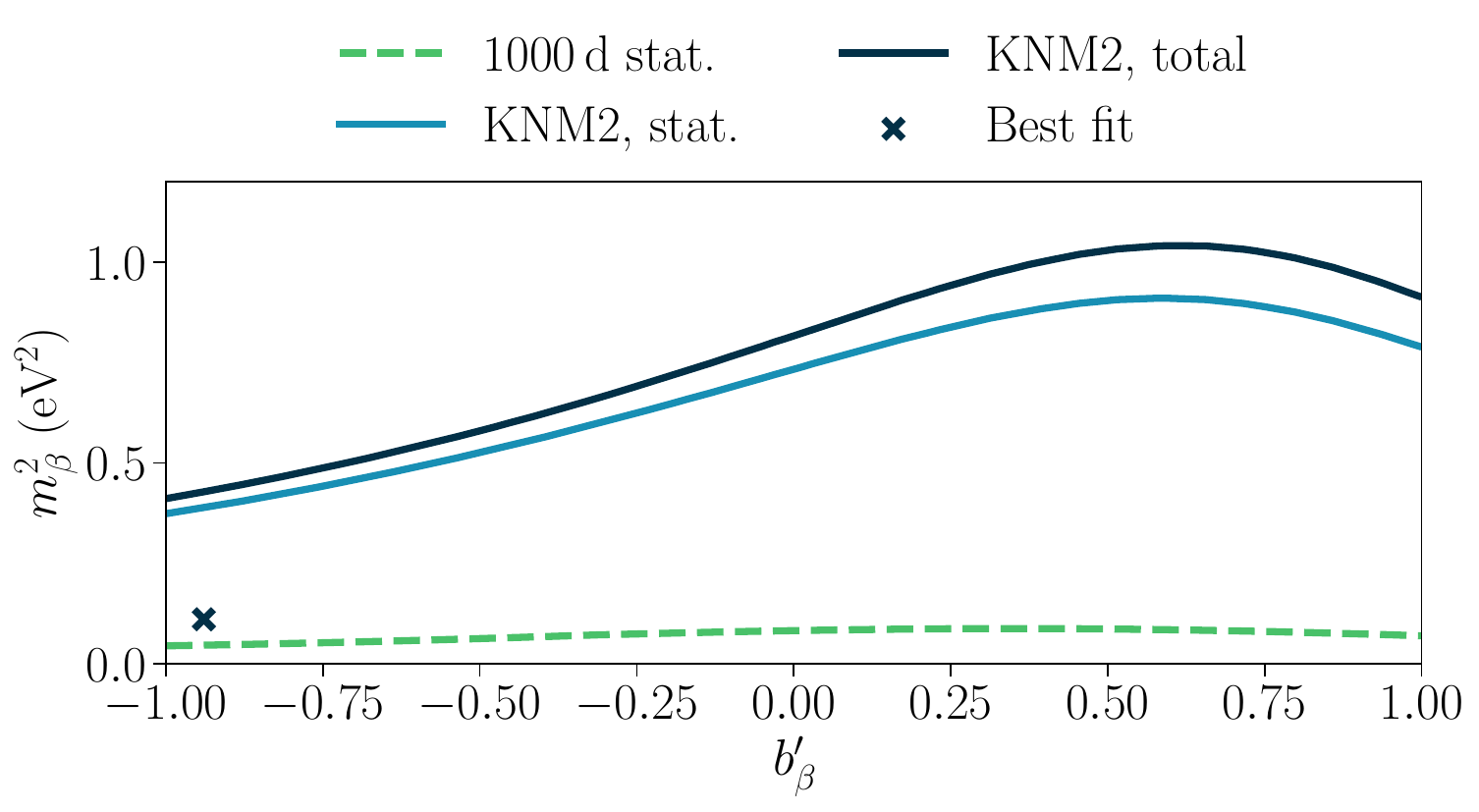}
    \caption{Exclusion contour for GNI contributions on the $\upbeta$-spectrum at 95\,\% CL based on the KNM2 data set. The contour is shown considering statistics only (stat.) and full systematics (total). The projected final sensitivity based on an Asimov data set with \num{1000} days statistics and $m_\upbeta^2=\SI{0}{eV^2}$ is shown by the green dashed line.}
    \label{fig:SM-contour}
\end{figure}

\begin{figure}
    \centering
    \includegraphics[width=.5\textwidth]{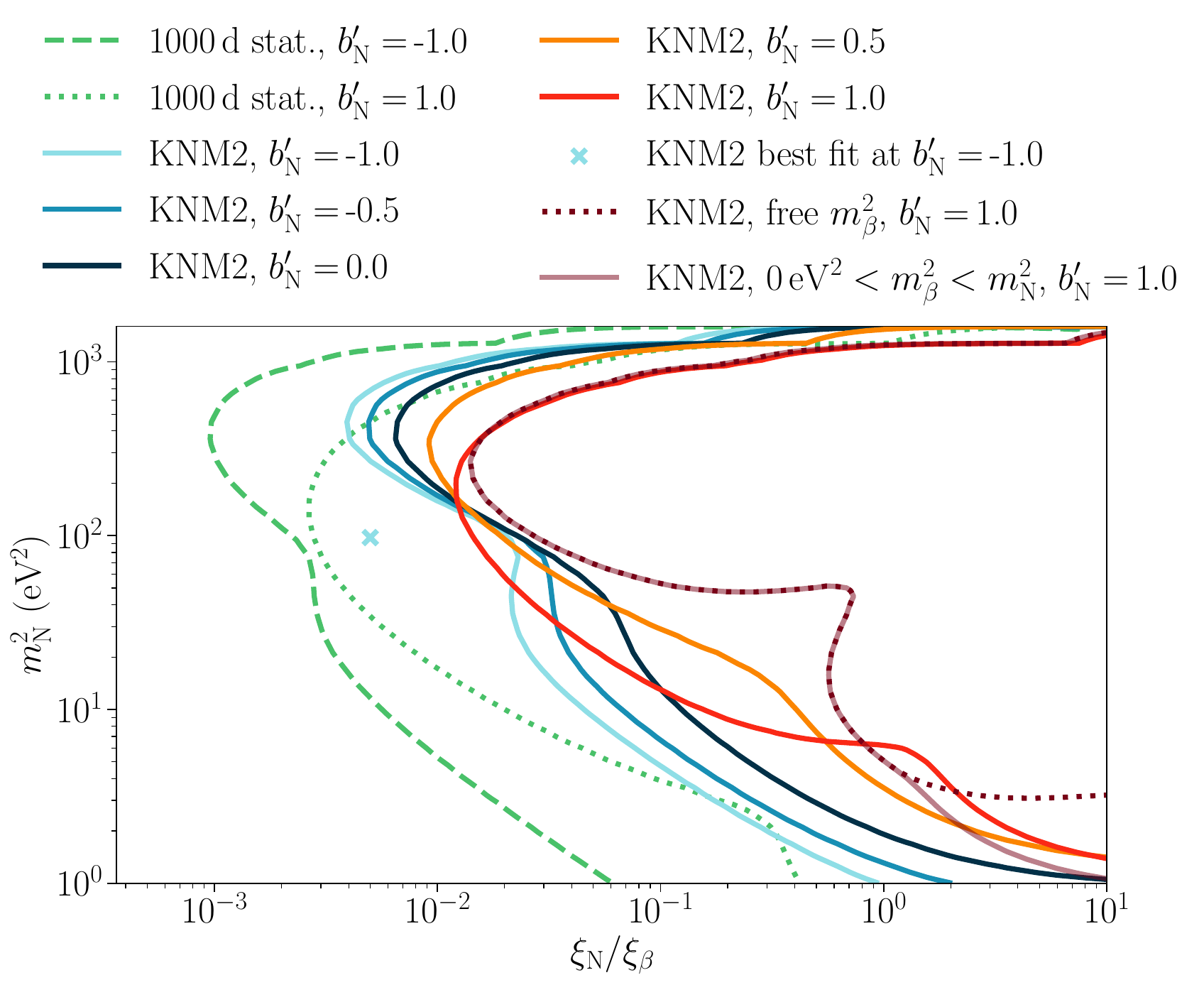}
    \caption{Exclusion contours for GNI contributions on an additional heavier neutrino-mass state at 95\,\% CL based on the KNM2 data set, taking into account full systematics. The two alternative treatments to fixing the light neutrino mass to $m_\upbeta^2=\SI{0}{eV^2}$ lead to the dark red contours. 
    The projected final sensitivity for two different values of $b'_\mathrm{N}$ based on an Asimov data set with \num{1000} days statistics and $m_\upbeta^2=\SI{0}{eV^2}$ is shown by the dashed and dotted green lines.}
    \label{fig:ST-contour}
\end{figure}

An interplay between the spectral components attributed to the light and heavy neutrino-mass states can occur during the fitting procedure due to correlations between the fitted masses of light and heavy neutrinos. 
When treating $m_\upbeta^2$ as a free parameter, a loss in sensitivity for both high and low values of $m_\mathrm{N}^2$ is observed.
To avoid unphysically large negative $m_\upbeta^2$ values, a different constraint $\SI{0}{eV^2}<m_\upbeta^2<m_\mathrm{N}^2$ is also used.
 The resulting exclusion contour shown in figure~\ref{fig:ST-contour} resembles the one using an unconstrained $m_\upbeta^2$ in the high-mass region, while regaining the sensitivity of the contour for fixed $m_\upbeta^2$ at low masses.\\

\textit{Constraints on GNI Couplings} \----
The presented results can be placed into the context of other experiments by comparing the limits on individual types of GNI via their $\epsilon$ couplings. This is necessary since each of the different experimental approaches observes a different combination of $\epsilon$ couplings. Here, we compare our results to experiments utilizing lepton-flavor-conserving charged-current interactions, such as nuclear and neutron $\upbeta$-decay, and pion decay. 
A more comprehensive comparison between experiments within the neutrino sector would only be possible under additional assumptions regarding the coupling behavior of the GNI. For instance, one might assume that the coupling strength is independent of the type of fermion. Furthermore, a comparison to high-energy processes is possible via the matching relations of the low-energy GNI terms to the SMNEFT Wilson coefficients.

The following limits on the individual contributions of the $\epsilon_i$ and $\tilde\epsilon_i$ terms can be derived from equations~\ref{equ:LudlFengler:Xi} to \ref{equ:LudlFengler:Xic} in the appendix. The values of the nuclear form factors are presented in table~\ref{tab:couplingconst}. When only considering light neutrinos we assume a neutrino mass of $m_\upbeta^2=\SI{1}{eV^2}$ in figure~\ref{fig:SM-contour} to provide an indication of the sensitivity of our data set to the magnitude of the individual $\epsilon_i$ contributions in a neutrino-mass regime in which this data set is sensitive to. We derive the following illustrative boundaries: $\operatorname{Re}(T_{\mathrm{e}i}\tilde\epsilon_\mathrm{L})>\num{-0.349}$, $\num{0.213}<\operatorname{Re}(T_{\mathrm{e}i}\tilde\epsilon_\mathrm{R})<\num{0.530}$, $\operatorname{Re}(T_{\mathrm{e}i}\tilde\epsilon_\mathrm{S})>\num{1.952}$, and $\num{-0.227}<\operatorname{Re}(T_{\mathrm{e}i}\tilde\epsilon_\mathrm{T})<\num{-0.083}$.  In the case of a heavy neutrino, our constraints on $\tilde\epsilon_i$ are derived from our high-sensitivity region at $m_\mathrm{N}^2=\SI{400}{eV^2}$ and summarized in table~\ref{tab:limits}. For all calculations we assume $|U_{\mathrm{e}i}|=|V_{\mathrm{e}i}|\approx 1$. Constraints on $\epsilon_i$ would become accessible for a left-right-symmetric coupling behavior of the GNI and a reasonable assumption on $T_{\mathrm{e}i}$ and $S_{\mathrm{e}i}$, respectively.

Given that the impact of the GNI on the $\upbeta$ spectrum scales with the mass of the neutrino, our investigations yield the most stringent constraints when including a heavier neutrino N. Furthermore, the strong correlations between $m_\upbeta$ and $b'_\upbeta$ visible in figure~\ref{fig:SM-contour}, and between $\tilde\epsilon_i$ and $T_{\mathrm{e}i}$ visible in equations~\ref{equ:LudlFengler:Xi} to \ref{equ:LudlFengler:Xic} in the appendix present a challenge in deriving competitive constraints on single $\epsilon_i$ contributions in comparison to the existing limits summarized in table~\ref{tab:limits}. 
When compared with other constraints from $\upbeta$-decay, as summarized in table~\ref{tab:limits}, it is demonstrated that competitive bounds on $\tilde\epsilon_i$ are derived for the left-handed vector and tensor interactions from only 5\,\% of the expected final KATRIN data set. The bound for the scalar interaction is less than one order of magnitude weaker than the competing constraints from $\upbeta$-decay experiments. 
The projected final sensitivity will enable a further improvement in these limits by about a factor of three.
High-energy investigations are overall one order of magnitude more sensitive to the $\tilde\epsilon_i$ couplings compared to the $\upbeta$-decay experiments and their bounds will come into reach for the full KATRIN data set. Also,  the impact of the markedly disparate energy scales and the associated running of the couplings has not been accounted for and should be kept in mind \cite{Bischer2019GeneralPerspective}.\\

\begin{table}[]
    \centering
    \caption{Overview of 90\,\% CL bounds on the GNI coefficients $\epsilon_{i,\,\mathrm{ud}}^{\mathrm{ee}11}$ and $\tilde\epsilon_{i,\,\mathrm{ud}}^{\mathrm{ee}11}$ from various low- and high-energy searches. MET stands for missing transverse energy of processes recorded at LHC. Limits derived for the SMNEFT Wilson coefficients can be associated with GNI coefficients and have been calculated assuming high-energy new physics (HNP) based on relations and limits given in \cite{Han2020ScalarInteractions}. The bounds derived in this work assume $m_\mathrm{N}^2=\SI{400}{eV^2}$.}
    \begin{tabular}{llc}\hline
      Bound (90\,\% CL) & Source & Reference \\\hline
       $\operatorname{Re}(\epsilon_\mathrm{L})<\num[exponent-product=\ensuremath{\cdot}]{5e-4}$ & $\upbeta$-decay & \cite{Cirigliano2013BetaEra}\\
       $\operatorname{Re}(\epsilon_\mathrm{R})<\num[exponent-product=\ensuremath{\cdot}]{5e-4}$  & $\upbeta$-decay & \cite{Cirigliano2013BetaEra}\\
       $|\epsilon_\mathrm{S}|<\num[exponent-product=\ensuremath{\cdot}]{3.4e-5}$ & SMNEFT, HNP & \cite{Han2020ScalarInteractions}\\
       $|\epsilon_\mathrm{S}|<\num[exponent-product=\ensuremath{\cdot}]{5.8e-3}$   & LHC (MET) & \cite{Naviliat-Cuncic2013ProspectsEra}\\
       $\operatorname{Re}(\epsilon_\mathrm{T})<\num[exponent-product=\ensuremath{\cdot}]{1e-3}$   & $\upbeta$-decay & \cite{Cirigliano2013BetaEra}\\
       $|\epsilon_\mathrm{T}|<\num[exponent-product=\ensuremath{\cdot}]{2e-4}$ & SMNEFT, HNP& \cite{Han2020ScalarInteractions}\\\hline
       $\operatorname{Re}(\tilde\epsilon_\mathrm{L})<\num[exponent-product=\ensuremath{\cdot}]{0.06}$ & $\upbeta$-decay & \cite{Cirigliano2013BetaEra}\\
       $|\tilde\epsilon_\mathrm{L}|< \num{0.063}-\num{0.131} $ & KATRIN (95\,\% CL) & this work \\
       $|\tilde\epsilon_\mathrm{R}|<\num[exponent-product=\ensuremath{\cdot}]{2.2e-3}$  & LHC (MET) & \cite{Naviliat-Cuncic2013ProspectsEra}\\
       $|\tilde\epsilon_\mathrm{R}|< \num{0.063}-\num{0.131} $ & KATRIN (95\,\% CL) & this work \\
       $|\tilde\epsilon_\mathrm{S}|\leq\num[exponent-product=\ensuremath{\cdot}]{0.063}$  &$\upbeta$-decay, CKM unitarity& \cite{Gonzalez-Alonso2018NewDecay}\\
       $|\tilde\epsilon_\mathrm{S}|<\num[exponent-product=\ensuremath{\cdot}]{5.8e-3}$   & LHC (MET)  & \cite{Naviliat-Cuncic2013ProspectsEra}\\
       $|\tilde\epsilon_\mathrm{S}|< \num{0.149}-\num{0.310} $ & KATRIN (95\,\% CL) & this work \\
       $\num[exponent-product=\ensuremath{\cdot}]{6e-4}\leq|\tilde\epsilon_\mathrm{T}|\leq\num{0.024}$ &$\upbeta$-decay& \cite{Gonzalez-Alonso2018NewDecay}\\
       $|\tilde\epsilon_\mathrm{T}|<\num[exponent-product=\ensuremath{\cdot}]{1.3e-3}$   & LHC (MET)  & \cite{Naviliat-Cuncic2013ProspectsEra}\\
       $|\tilde\epsilon_\mathrm{T}|< \num{0.022}-\num{0.046} $ & KATRIN (95\,\% CL) & this work \\\hline
    \end{tabular}
    \label{tab:limits}
\end{table}

\begin{figure}
    \centering
    \includegraphics[width=.5\textwidth]{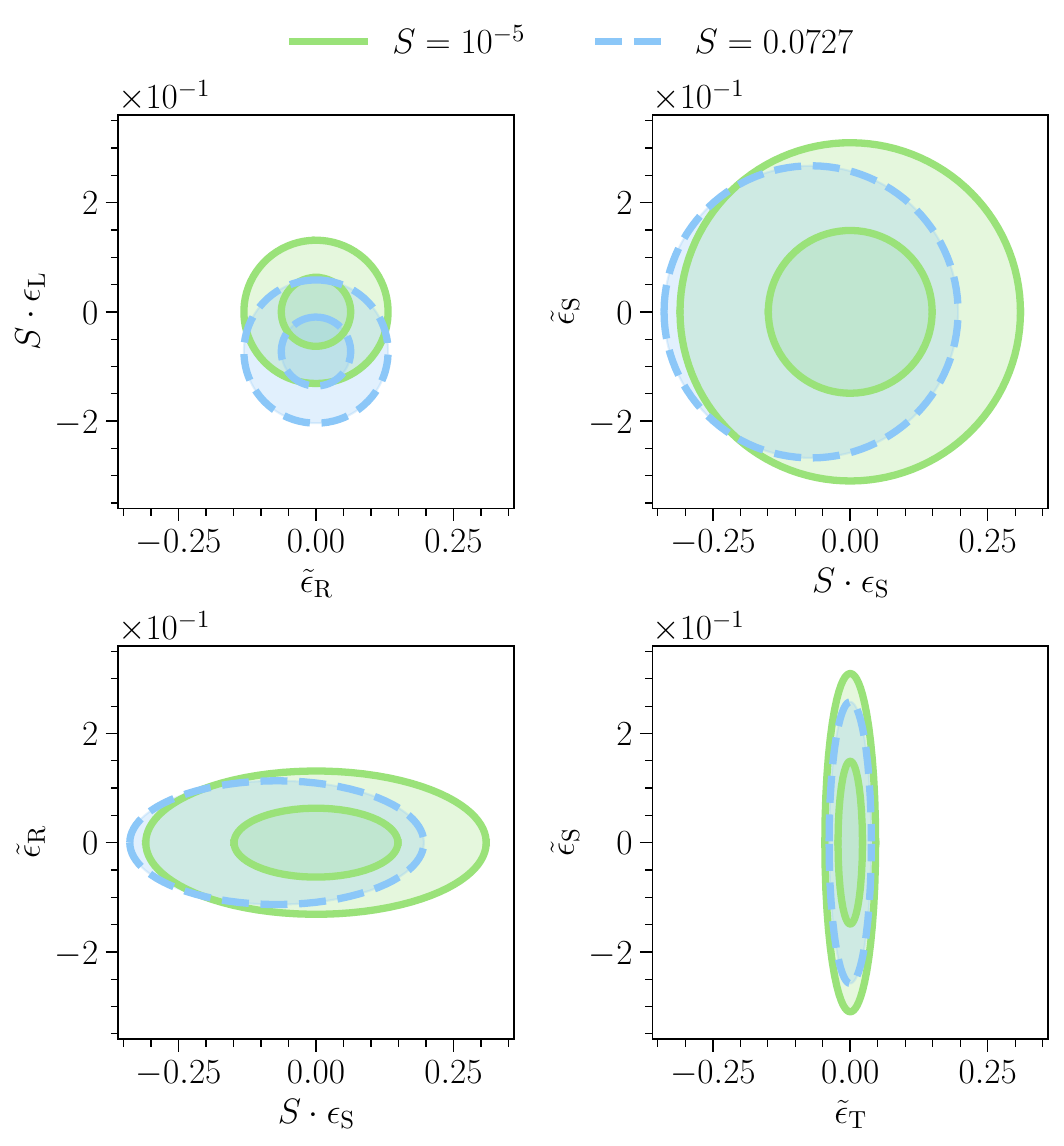}
    \caption{95\,\% CL exclusion contours for selected combinations of $\epsilon_i$ and $\tilde\epsilon_i$ derived from the exclusion contours on KNM2 data assuming $m_\upbeta^2=\SI{0}{eV^2}$
    for the high-sensitivity region at $m_\mathrm{N}^2=\SI{400}{eV^2}$. Colored regions display the not excluded region in the range of $b'_\mathrm{N}$ between -1 and 1, and depend on the choice of mixing parameter S. If parameters run out of their defined range, only the contour for $b'_\mathrm{N}=1$ is shown.} 
    \label{fig:eps-contour}
\end{figure}

Furthermore, we examine simultaneous constraints on more than one interaction type. In this study, we have selected combinations that may provide indications of the existence of specific new-physics scenarios. The resulting contours are presented in figure~\ref{fig:eps-contour} for $m_\mathrm{N}^2=\SI{400}{eV^2}$ and $b'_\mathrm{N}$ between -1 and 1. Two values for the unknown mixing $S$ are explored: an external, conservative constraint of $S=0.0727$ from $0\nu\upbeta\upbeta$-decay \cite{Aker2022ImprovedCampaign,Jana2024RestrictingDecay} and a small mixing of $S=\num{e-5}$.

The W boson, which mediates the $\upbeta$-decay, only couples to left-handed fermions. Incorporating right-handed neutrino states into our theoretical framework enables the existence of a right-handed vector-current interaction, which may be mediated by a right-handed W boson. This scenario would thus necessitate both $\epsilon_\mathrm{L}$ and $\tilde\epsilon_\mathrm{R}$ to be non-zero. Both $S\cdot\epsilon_\mathrm{L}$ and $\tilde\epsilon_\mathrm{R}$ can be simultaneously constrained at the order of $\num{e-1}$. 
Furthermore, the GNI introduce scalar charged-current interactions, represented by $\epsilon_\mathrm{S}$ and $\tilde\epsilon_\mathrm{S}$, which can indicate the existence of a charged Higgs. The exclusion contours in figure~\ref{fig:eps-contour} demonstrate that if such a charged scalar interaction couples both to left- and right-handed neutrinos, the bounds on both $S\cdot\epsilon_\mathrm{S}$ and $\tilde\epsilon_\mathrm{S}$ are below 0.35.
Additionally, we present the constraints on the combination of the couplings $\epsilon_\mathrm{S}$ and $\tilde\epsilon_\mathrm{R}$, which may provide insights into the combined interaction of a charged Higgs acting on the left-handed neutrino state and a right-handed W boson. The constraints are at an $\num{e-1}$ level for $\tilde\epsilon_\mathrm{R}$ while encountering a factor of 2 lower sensitivity in the $S\cdot\epsilon_\mathrm{S}$ dimension.
Finally, the matching relations in \cite{Bischer2019GeneralPerspective} enable us to derive which interactions would be needed to realize various leptoquark models. These would serve as a unified approach to UV completion of the effective GNI theory \cite{Bischer2019GeneralPerspective}. As an illustrative example we present the combination of the couplings $\tilde\epsilon_\mathrm{S}$ and $\tilde\epsilon_\mathrm{T}$ for the scalar color-triplet leptoquark $R'_2$, as defined in \cite{Bischer2019GeneralPerspective}. We constrain $\tilde\epsilon_\mathrm{S}$ to a precision of below \num{5e-2} while the constraint on $\tilde\epsilon_\mathrm{T}$ is about half an order of magnitude larger. The combination with one other scalar color-triplet leptoquark $S_1$ is of interest as it allows for the explanation of B physics anomalies \cite{Cata2019LinkingAnomalies}, and is able to generate radiative neutrino masses at one-loop level \cite{Klein2019MinimalMasses}. The above investigated combinations, namely $\epsilon_\mathrm{L}$ and $\tilde\epsilon_\mathrm{R}$, $\epsilon_\mathrm{S}$ and $\tilde\epsilon_\mathrm{S}$, and $\tilde\epsilon_\mathrm{S}$ and $\tilde\epsilon_\mathrm{T}$, are expected to occur in the presence of the $S_1$ leptoquark.\\

\textit{Conclusions and Outlook} \----
This work presents the first search for general neutrino interactions at the KATRIN experiment. We derived an effective description of general neutrino interactions in tritium $\upbeta$-decay and applied this model to the data of the second KATRIN measurement campaign. In these investigations, the possibility of the existence of an additional heavier neutrino with $m_\mathrm{N}^2\leq\SI{1600}{eV^2}$ was considered. In the absence of a significant signal for general neutrino interactions, we present the exclusion contours at 95\,\% CL on the effective GNI parameters.

The impact of systematic effects was studied in detail for both the case with and without the inclusion of an additional heavier neutrino in our model. 
We have shown that our sensitivity is clearly statistics-limited for the KNM2 data set.

We derived constraints on the coupling strength of single types of interactions on left- and right-handed neutrinos, designated as $\epsilon_i$ and $\tilde\epsilon_i$, respectively. We are able to place competitive constraints with respect to global analyses on $\upbeta$-decay processes for left-handed vector and tensor interactions for high masses of an additional neutrino with $m_\mathrm{N}^2=\SI{400}{eV^2}$. 

Furthermore, we investigated more specific new physics scenarios, including a right-handed W boson, charged Higgs boson, and scalar leptoquark, by drawing exclusion contours for selected combinations of the coupling strengths, $\epsilon_i$ and $\tilde\epsilon_i$.

With the full data set of KATRIN collected until 2025, a significant threefold enhancement in GNI sensitivity is expected, which would entail setting leading constraints among $\upbeta$-decay experiments.
As the sensitivity to the effective GNI parameters scales with the neutrino mass, the planned detector update TRISTAN, designed to search for keV-scale sterile neutrinos, promises significant future potential for further GNI searches.\\


\begin{acknowledgments}
The authors thank Patrick Otto Ludl, Thomas Schwetz-Mangold, and Andreas Trautner for the helpful exchange on theory.
We acknowledge the support of Helmholtz Association (HGF), Ministry for Education and Research BMBF (05A23PMA, 05A23PX2, 05A23VK2, and 05A23WO6), the doctoral school KSETA at KIT, Helmholtz Initiative and Networking Fund (grant agreement W2/W3-118), Max Planck Research Group (MaxPlanck@TUM), and Deutsche Forschungsgemeinschaft DFG (GRK 2149 and SFB-1258 and under Germany's Excellence Strategy EXC 2094 – 390783311) in Germany; Ministry of Education, Youth and Sport (CANAM-LM2015056, LTT19005) in the Czech Republic; Istituto Nazionale di Fisica Nucleare (INFN) in Italy; the National Science, Research and Innovation Fund via the Program Management Unit for Human Resources \& Institutional Development, Research and Innovation (grant B37G660014) in Thailand; and the Department of Energy through grants DE-FG02-97ER41020, DE-FG02-94ER40818, DE-SC0004036, DE-FG02-97ER41033, DE-FG02-97ER41041, {DE-SC0011091 and DE-SC0019304 and the Federal Prime Agreement DE-AC02-05CH11231} in the United States. This project has received funding from the European Research Council (ERC) under the European Union Horizon 2020 research and innovation programme (grant agreement No. 852845). We thank the computing cluster support at the Institute for Astroparticle Physics at Karlsruhe Institute of Technology, Max Planck Computing and Data Facility (MPCDF), and the National Energy Research Scientific Computing Center (NERSC) at Lawrence Berkeley National Laboratory.
\end{acknowledgments}

\appendix
\section{Appendix}
\textit{Derivation of an Effective Description of GNI for Tritium $\upbeta$-Decay} \----
\begin{table}
    \caption{Definition of general neutrino interaction coupling constants and operators appearing in $\mathcal{L}^\mathrm{CC}_\mathrm{GNI}$ in equation~\ref{equ:CC}.}
\begin{ruledtabular}
    \centering
    \begin{tabular}{cccc}
      $j$   & $\epsilon_j$ & $\mathcal{O}_j$ & $\mathcal{O}'_j$ \\\hline
      1 & $\epsilon_\mathrm{L}$ & $\gamma_{\mu}(\mathds{1}-\gamma^5)$ & $\gamma^{\mu}(\mathds{1}-\gamma^5)$ \\
      2 & $\tilde{\epsilon}_\mathrm{L}$ & $\gamma_{\mu}(\mathds{1}+\gamma^5)$ & $\gamma^{\mu}(\mathds{1}-\gamma^5)$ \\
      3 & $\epsilon_\mathrm{R}$ & $\gamma_{\mu}(\mathds{1}-\gamma^5)$ & $\gamma^{\mu}(\mathds{1}+\gamma^5)$ \\
      4 & $\tilde{\epsilon}_\mathrm{R}$ & $\gamma_{\mu}(\mathds{1}+\gamma^5)$ & $\gamma^{\mu}(\mathds{1}+\gamma^5)$ \\
      5 & $\epsilon_\mathrm{S}$ & $(\mathds{1}-\gamma^5)$ & $\mathds{1}$ \\
      6 & $\tilde{\epsilon}_\mathrm{S}$ & $(\mathds{1}+\gamma^5)$ & $\mathds{1}$ \\
      7 & $-\epsilon_\mathrm{P}$ & $(\mathds{1}-\gamma^5)$ & $\gamma^5$ \\
      8 & $-\tilde{\epsilon}_\mathrm{P}$ & $(\mathds{1}+\gamma^5)$ & $\gamma^5$ \\
      9 & $\epsilon_\mathrm{T}$ & $\sigma_{\mu\nu}(\mathds{1}-\gamma^5)$ & $\sigma^{\mu\nu}(\mathds{1}-\gamma^5)$ \\
     10 & $\tilde{\epsilon}_\mathrm{T}$ & $\sigma_{\mu\nu}(\mathds{1}+\gamma^5)$ & $\sigma^{\mu\nu}(\mathds{1}+\gamma^5)$ \\
    \end{tabular}
    \label{tab:GNIOperator}
\end{ruledtabular}
\end{table}
For deriving the effective $\upbeta$-decay rate under the influence of GNI, reference \cite{Ludl2016DirectInteractions} is taken as a starting point.
The effective Lagrangian used for this derivation is composed of two elements: the Lagrangian describing the Standard Model contribution and the charged-current GNI Lagrangian, denoted by $\mathcal{L}^\mathrm{CC}_\mathrm{GNI}$ and defined in equation~\ref{equ:CC}. The GNI operators are defined in table~\ref{tab:GNIOperator}.
In order to obtain the squared matrix element of the decay, $|\mathcal{M}|^2$, a sum of traces over all possible combinations of GNI and Standard Model interactions must be calculated. Integration of $|\mathcal{M}|^2$ over the neutrino energy provides a complete description of the total differential decay rate. To simplify the lengthy new description of the total differential decay rate to be applicable for a search for new physics on the $\upbeta$-spectrum, we expand in small parameters \cite{Ludl2016DirectInteractions}.
To incorporate nuclear effects into the quark-level calculations, nuclear matrix elements are taken into account.
The relevant matrix elements for tritium beta decay can be found in \cite{Cirigliano2013BetaEra}. Given that $q\equiv p_n-p_p <\SI{20}{keV}$ and $M_\mathrm{N}\equiv (m_\mathcal{A}+m_\mathcal{B})/2\simeq\SI{3}{GeV}$ for tritium decay, it is sufficient to consider terms up to $\mathcal{O}(q/M_\mathrm{N})<\num{e-5}$ in the calculation. The matrix element associated with the pseudoscalar coupling is suppressed at the order \num{e-5} and therefore not further considered due to its negligible impact.
Also, the contribution of the weak magnetism is assumed to be negligible. 
\begin{table}
    \caption{Numerical values for the nuclear form factor used in the calculations at hand. (CVC = conserved vector current).} 
    \label{tab:couplingconst}
\begin{ruledtabular}
    \centering
    \begin{tabular}{ccl}
    Coupling & Value & Reference, Comments   \\\hline
       $g_\mathrm{V}$  & 1 & \cite{Simkovic2007ExactModel}, (assuming CVC) $^3$H \\
        $\frac{g_\mathrm{A}}{g_\mathrm{V}}$ & $1.2646 \pm 0.0035$ & \cite{Akulov2002DeterminationDecay}, $^3$H  \\
        $g_\mathrm{S}$ & $1.02 \pm 0.11$ & \cite{Gonzalez-Alonso2014IsospinPhysics}, $\overline{\text{MS}}$, neutron \\
        $g_\mathrm{P}$ & $349 \pm 9$ & \cite{Gonzalez-Alonso2014IsospinPhysics}, $\overline{\text{MS}}$, neutron \\
        $g_\mathrm{T}$ & $0.987 \pm 0.051 \pm 0.020$ & \cite{Bhattacharya2016AxialQCD}, lattice, $\overline{\text{MS}}$, neutron \\
    \end{tabular}
\end{ruledtabular}
\end{table}
This leads to an effective description of the GNI contributions in terms of the GNI parameters $\xi$, $b$, $b'$ and $c$ as

    \begin{eqnarray}\nonumber
    &\left(\frac{\mathrm{d}\Gamma}{\mathrm{d}E}\right)_j=\frac{G_\mathrm{F}^2 V_\mathrm{ud}^2}{2\pi^3}\sqrt{(E+m_\mathrm{e})^2-m_\mathrm{e}^2}\\\nonumber
    &\cdot (E+m_\mathrm{e})(E_0-E)
    \sqrt{(E_0-E)^2-m_j^2}\\
    &\cdot \xi \left[1+b\frac{m_\mathrm{e}}{E+m_\mathrm{e}}-b'\frac{m_j}{E_0-E}-c\frac{m_\mathrm{e} m_j}{(E+m_\mathrm{e})(E_0-E)}\right],
\end{eqnarray}
where $m_j$ are the individual neutrino-mass states. In this analysis, we employ an effective mass description, wherein the neutrino masses $m_j$ are considered to be fully degenerate. This is a valid approximation, as the individual neutrino-mass states cannot be resolved by the current setup. Consequently, the sum over the neutrino-mass states can be treated as the effective neutrino mass, $m_\upbeta$. Since the current setup is not sensitive to the parameters $b$ or to $b'$ and $c$ individually, we apply $E\ll m_\mathrm{e}$ in the analysis. The corresponding redefinition is $\xi_k+\xi_k b_k \rightarrow \xi_k$ and $\xi_k b'_k + \xi_k c_k \rightarrow\xi_k b'_k$. For future reference and better comparison with published results, we give the explicit definitions for the four GNI parameters separately in equations~\ref{equ:LudlFengler:Xi} to \ref{equ:LudlFengler:Xic}.

The analysis distinguishes between a scenario with (N) and without ($\upbeta$) an additional heavier neutrino-mass state. 
In accordance to \cite{Ludl2016DirectInteractions}, we define $m_i$ and $M_j$ to be the masses of the light and heavy neutrino-mass eigenstates $\nu_i'$ and $\mathrm{N}_{\mathrm{R}j}'$, respectively. The left- and right-handed neutrino flavor fields with the mixing matrices $U$, $T$, $S$, and $V$ are given by
\begin{eqnarray}
    \nu_\mathrm{L}&=&U\,\nu_\mathrm{L}' + S\, \mathrm{N}_\mathrm{R}^{\prime\;c},\\
    \nu_\mathrm{R}&=&T^\ast\,\nu_\mathrm{L}^{\prime\;c} + V^\ast\, \mathrm{N}'_\mathrm{R}.
\end{eqnarray}
$S$ and $T$ are presumably small. The explicit dependency of the GNI parameters is analogous for both cases. Therefore, we present the results only for the light neutrino-mass state. With $U \rightarrow S$ and $T \rightarrow V$ the results are valid for the heavy neutrino-mass state.

\begin{eqnarray}\label{equ:LudlFengler:Xi}
\nonumber
       & \xi_\upbeta =|U_{\mathrm{e}i}|^2 \bigl[ (g_\mathrm{V}^2+3g_\mathrm{A}^2)(1+|\epsilon_\mathrm{L}|^2+|\epsilon_\mathrm{R}|^2+2\operatorname{Re}(\epsilon_\mathrm{L}))\\\nonumber
    &+2(g_\mathrm{V}^2-3g_\mathrm{A}^2)(\operatorname{Re}(\epsilon_\mathrm{R}\epsilon^\ast_\mathrm{L})+\operatorname{Re}(\epsilon_\mathrm{R})) 
    +g_\mathrm{S}^2 |\epsilon_\mathrm{S}|^2\\\nonumber
    &+48g_\mathrm{T}^2|\epsilon_\mathrm{T}|^2\bigr]
     +|T_{\mathrm{e}i}|^2 \bigl[ (g_\mathrm{V}^2+3g_\mathrm{A}^2)(|\tilde\epsilon_\mathrm{L}|^2+|\tilde\epsilon_\mathrm{R}|^2)\\
    &+2(g_\mathrm{V}^2-3g_\mathrm{A}^2)\operatorname{Re}(\tilde\epsilon_\mathrm{R}\tilde\epsilon^\ast_\mathrm{L})
    +g_\mathrm{S}^2 |\tilde\epsilon_\mathrm{S}|^2
    +48g_\mathrm{T}^2|\tilde\epsilon_\mathrm{T}|^2\bigr]
\end{eqnarray}

\begin{eqnarray}\label{equ:LudlFengler:Xib}\nonumber
     & \xi_\upbeta b_\upbeta= |U_{\mathrm{e}i}|^2 \bigl[2 g_\mathrm{S}g_\mathrm{V}\operatorname{Re}(\epsilon_\mathrm{S}(1+\epsilon_\mathrm{L}+\epsilon_\mathrm{R})^\ast)\\\nonumber
    &-24g_\mathrm{A}g_\mathrm{T}\operatorname{Re}(\epsilon_\mathrm{T}(1+\epsilon_\mathrm{L}-\epsilon_\mathrm{R})^\ast)\bigr]
    +|T_{\mathrm{e}i}|^2 \bigl[2 g_\mathrm{S}g_\mathrm{V}\\
    &\cdot\operatorname{Re}(\tilde\epsilon_\mathrm{S}(\tilde\epsilon_\mathrm{L}+\tilde\epsilon_\mathrm{R})^\ast)
    +24g_\mathrm{A}g_\mathrm{T}\operatorname{Re}(\tilde\epsilon_\mathrm{T}(\tilde\epsilon_\mathrm{L}-\tilde\epsilon_\mathrm{R})^\ast)\bigr]
\end{eqnarray}

\begin{eqnarray}\label{equ:LudlFengler:Xib'}\nonumber
        &\xi_\upbeta b'_\upbeta = \operatorname{Re}(U_{\mathrm{e}i}T_{\mathrm{e}i}) \bigl[ 2 g_\mathrm{V} g_\mathrm{S} (\operatorname{Re}(\tilde\epsilon_\mathrm{S}(1+\epsilon_\mathrm{L}+\epsilon_\mathrm{R})^\ast)\\\nonumber
        &+\operatorname{Re}(\epsilon_\mathrm{S}(\tilde\epsilon_\mathrm{L}+\tilde\epsilon_\mathrm{R})^\ast))
        -24 g_\mathrm{A} g_\mathrm{T} (\operatorname{Re}(\tilde\epsilon_\mathrm{T}\\
        &\cdot (1+\epsilon_\mathrm{L}-\epsilon_\mathrm{R})^\ast)
        -\operatorname{Re}(\epsilon_\mathrm{T}(\tilde\epsilon_\mathrm{L}-\tilde\epsilon_\mathrm{R})^\ast)) \bigr]
\end{eqnarray}

\begin{eqnarray}\label{equ:LudlFengler:Xic}\nonumber
    &\xi_\upbeta c_\upbeta = \operatorname{Re}(U_{\mathrm{e}i}T_{\mathrm{e}i}) \bigl[ 2 g_\mathrm{V}^2\operatorname{Re}((\tilde\epsilon_\mathrm{L}+\tilde\epsilon_\mathrm{R})(1+\epsilon_\mathrm{L}+\epsilon_\mathrm{R})^\ast)\\\nonumber
        & -6 g_\mathrm{A}^2\operatorname{Re}((\tilde\epsilon_\mathrm{L}-\tilde\epsilon_\mathrm{R})(1+\epsilon_\mathrm{L}-\epsilon_\mathrm{R})^\ast)\\
        & +2 g_\mathrm{S}^2 \operatorname{Re}(\epsilon_\mathrm{S}\tilde\epsilon_\mathrm{S}^\ast)
         + 96 g_\mathrm{T}^2 \operatorname{Re}(\epsilon_\mathrm{T}\tilde\epsilon_\mathrm{T}^\ast)
\end{eqnarray}
These equations agree with the corresponding formulas in references \cite{Enz1957FermiParity, Gluck1995MeasurableDecay,Bischer2021EffectivePhenomenology,Bonn2011TheDecay}. In order to derive the limits on the $\epsilon_i$ and $\tilde\epsilon_i$ couplings, we assume the above-defined terms to be real.\\

\begin{figure}
    \centering
    \includegraphics[width=.49\textwidth]{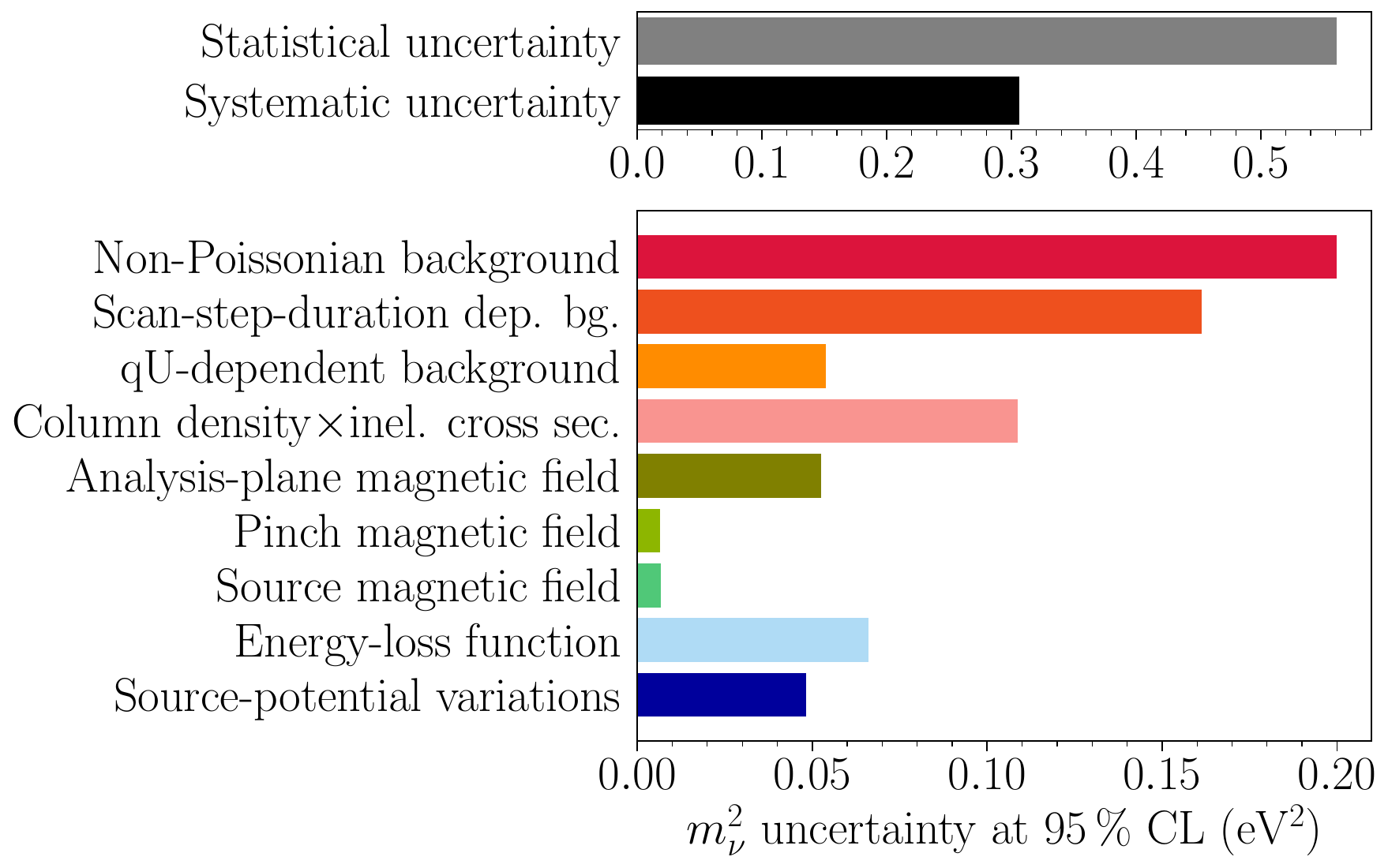}
    \caption{Uncertainty breakdown showing the impact of the KNM2 systematic effects on the GNI sensitivity contour based on the $\upbeta$-spectrum. The breakdown is shown for $b'_\upbeta=0$ and obtained from a simulated data set without fluctuations with $m_\upbeta^2=\SI{0}{eV^2}$. The uncertainty is dominated by its statistical component. The most significant systematic effects are the non-Poissonian and the scan-step-duration dependent background.}
    \label{fig:SysBreaksownSM}
\end{figure}

\begin{figure}
    \centering
    \includegraphics[width=.49\textwidth]{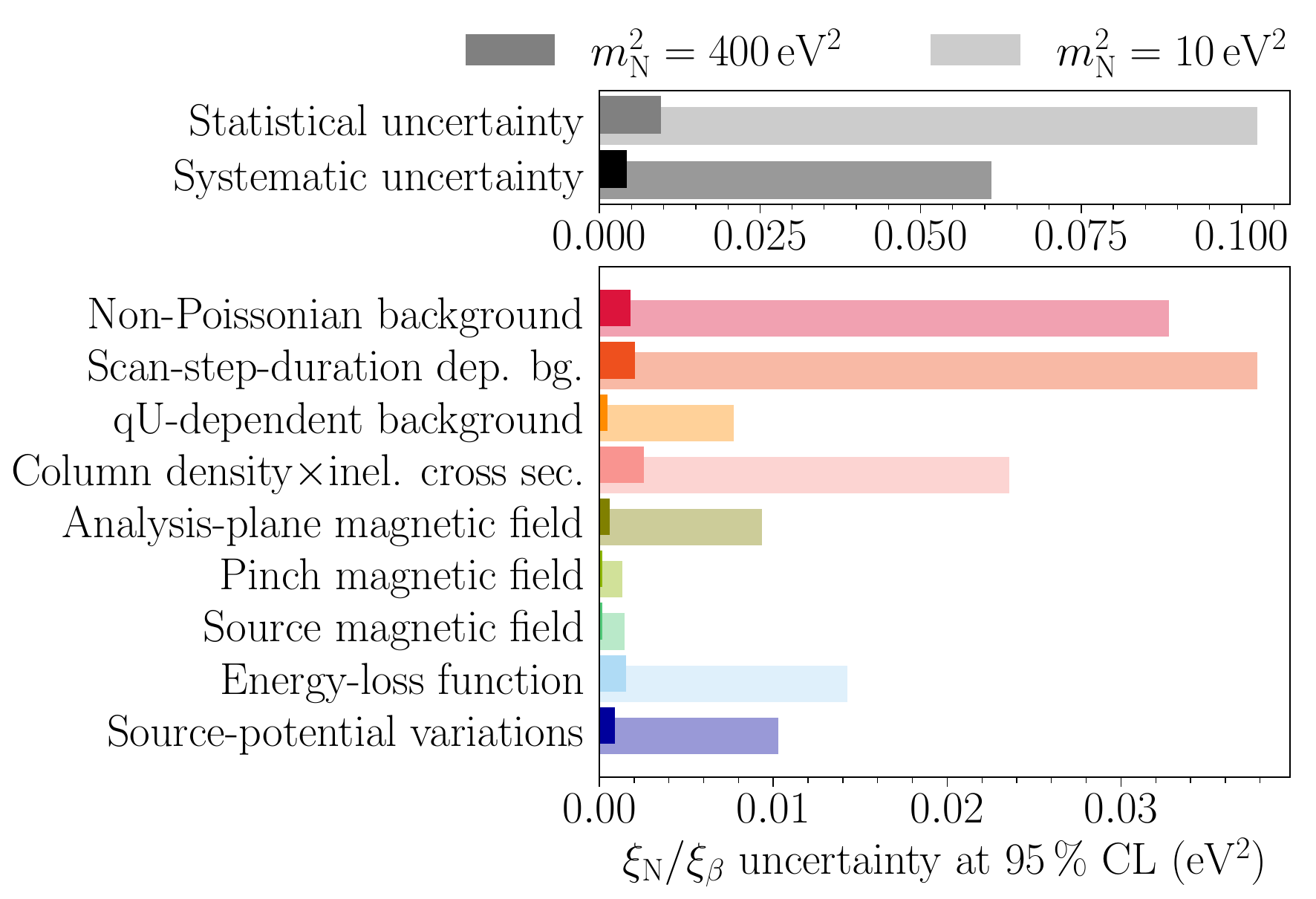}
    \caption{Uncertainty breakdown showing the impact of the KNM2 systematic effects on the GNI sensitivity contour taking into account the presence of an additional heavier neutrino (N). The breakdown is shown for the lower- and higher-mass region of the additional mass state with $b'_\mathrm{N}=0$ and obtained from a simulated data set without fluctuations with $m_\upbeta^2=\SI{0}{eV^2}$. In both cases of $m_\mathrm{N}^2$ the uncertainty is statistics dominated. In the lower-mass region the most significant systematic effects are the non-Poissonian and scan-step-duration dependent background, while the energy-loss function and column density gain significance towards the higher-mass region.}
    \label{fig:SysBreakdownST}
\end{figure}

\textit{Systematic Uncertainties} \----
In the presented analysis 
all known systematic effects from the neutrino-mass analysis \cite{Aker2024DirectData} are considered. The breakdown of systematic uncertainties on the contours is performed on a simulated data set with the neutrino mass set to zero. Grid scans are performed, wherein each systematic effect is considered separately. 
The analysis shows that the sensitivity of the contour is limited by statistics while the systematic effects only cause slight alterations. 
To quantitatively assess the relative contribution of each individual systematic effect \cite{Aker2022DirectSensitivity}, raster scans are performed. For each value of $m_4^2$ and $b'_\upbeta$, respectively, the 95\,\% CL contour for one degree of freedom is drawn, and the squared difference with respect to the statistics-only contour is calculated to obtain the pure systematic contribution.
The impact of systematic effects on the GNI sensitivity is shown in figures \ref{fig:SysBreaksownSM} and \ref{fig:SysBreakdownST}.
The low signal rate for small neutrino masses $m_\upbeta$ and $m_\mathrm{N}$ gives relevance to the background-related systematics, such as the uncertainty on the non-Poissonian distributed background events and the scan-step-duration dependent background component coming from electrons stored in the Penning trap between the pre- and main spectrometers. Towards large values of $m_\mathrm{N}$ the signal-to-background ratio enlarges, reducing the relevance of the background systematics while highlighting the dominance of other systematics, such as the uncertainty contributions of the energy-loss function and the column density.

\bibliography{apssamp}
\end{document}